\documentclass[pre,tighten]{revtex4}

\usepackage{amssymb,amsmath}

\usepackage{graphicx}
\newcommand{\dd}{\text{d}}
\begin{document}
\title{Enskog kinetic theory for $d$-dimensional dense granular gases}
\author{Vicente Garz\'{o}}
\address{Departamento de F\'{\i}sica, Universidad de Extremadura, E-06071
Badajoz, Spain}
\email[E-mail: ]{vicenteg@unex.es}
\begin{abstract}
The goal of this note is to provide most of the technical details involved in the application of the Chapman-Enskog method to solve the revised Enskog equation to Navier-Stokes order. Explicit expressions for the transport coefficients and the cooling rate are obtained in terms of the coefficient of restitution and the solid volume fraction by using a new Sonine approach. This new approach consists of replacing, where appropriate in the Chapman-Enskog procedure, the local equilibrium distribution (used in the standard first Sonine approximation) by the homogeneous cooling state distribution. The calculations are performed in an arbitrary number of dimensions.
\end{abstract}

\draft

\date{\today}
\maketitle

\section{First-order distribution function}

We consider a granular fluid composed of smooth inelastic hard
spheres ($d=3$) or disks ($d=2$) of mass $m$ and diameter $\sigma$. Collisions are
characterized by a (constant) coefficient of normal restitution
$0<\alpha\leq 1$. At a kinetic level, all the relevant information
on the system is given through the one-particle velocity
distribution function $f({\bf r}, {\bf v}, t)$, which is assumed to obey the (inelastic) Enskog equation. This equation can be solved by the Chapman-Enskog method. To first order in the spatial gradients (Navier-Stokes order), the distribution function $f^{(1)}({\bf r}, {\bf v}, t)$ is given by \cite{GD99,L05}
\begin{eqnarray}
\label{1}
f^{(1)}&=&\boldsymbol{\mathcal{A}}\left(
\mathbf{V}\right)\cdot  \nabla \ln
T+\boldsymbol{\mathcal{B}}\left(
\mathbf{V}\right) \cdot \nabla \ln n
\nonumber\\
& & +\mathcal{C}_{ij}\left( \mathbf{V} \right)\frac{1}{2}\left( \partial _{i}U_{j}+\partial _{j
}U_{i}-\frac{2}{d}\delta _{ij}\nabla \cdot
\mathbf{U} \right)+\mathcal{D}\left( \mathbf{V} \right) \nabla \cdot
\mathbf{U}.
\end{eqnarray}
The quantities $\boldsymbol{\mathcal{A}}$, $\boldsymbol{\mathcal{B}}$, $\mathcal{C}_{ij}$ and $\mathcal{D}$ are the solutions of the following integral equations \cite{GD99}
\begin{equation}
\frac{1}{2}\zeta^{(0)}\frac{\partial}{\partial {\bf V}}\cdot \left({\bf V}\boldsymbol{\mathcal{A}}\right)-
\frac{1}{2}\zeta^{(0)}\boldsymbol{\mathcal{A}}+
\mathcal{L}\boldsymbol{\mathcal{A}}=\mathbf{A},  \label{2}
\end{equation}
\begin{equation}
\frac{1}{2}\zeta^{(0)}\frac{\partial}{\partial {\bf V}}\cdot \left({\bf V}\boldsymbol{\mathcal{B}}\right)+
\mathcal{L}\boldsymbol{\mathcal{B}}=\mathbf{B}+\zeta^{(0)}\left(1+\phi\frac{\partial}{\partial \phi}\ln \chi\right)\boldsymbol{\mathcal{A}},  \label{3}
\end{equation}
\begin{equation}
\frac{1}{2}\zeta^{(0)}\frac{\partial}{\partial {\bf V}}\cdot \left({\bf V}\mathcal{C}_{ij}\right)+\frac{1}{2}\zeta^{(0)}\mathcal{C}_{ij}+
\mathcal{L}\mathcal{C}_{ij}=C_{ij},  \label{4}
\end{equation}
\begin{equation}
\frac{1}{2}\zeta^{(0)}\frac{\partial}{\partial {\bf V}}\cdot \left({\bf V}\mathcal{D}\right)+\frac{1}{2}\zeta^{(0)}\mathcal{D}+
\mathcal{L}\mathcal{D}=D,  \label{5}
\end{equation}
where $\phi=(\Omega_d/2^d d)n\sigma^d$ is the solid volume fraction and $\Omega_d=2\pi^{d/2}/\Gamma(d/2)$ is the area of the $d$-dimensional unit hard sphere. The linear operator $\mathcal{L}$ is given by
\begin{equation}
\mathcal{L}X=-\left(J^{(0)}[f^{(0)},X]+J^{(0)}[X,f^{(0)}]\right),  \label{6}
\end{equation}
where the collision operator $J^{(0)}[X,Y]$ can be recognized as the Boltzmann operator for inelastic collisions multiplied by the pair correlation function $\chi(\phi)$. The inhomogeneous terms of Eqs.\ \eqref{2}--\eqref{5} are defined by
\begin{equation}
{\bf A}\left( \mathbf{V}\right)=\frac{1}{2} {\bf V}
\nabla_{\mathbf{V}}\cdot \mathbf{V}f^{(0)}
-\frac{p}{\rho }\nabla
_{\mathbf{V}}f^{(0)}+\frac{1}{2}\boldsymbol{\mathcal{K}}\left[\nabla_{\mathbf{V}}\cdot \left( \mathbf{V}
f^{(0)}\right) \right] ,  \label{7}
\end{equation}
\begin{equation}
{\bf B}\left(\mathbf{V}\right) = -{\bf V}f^{(0)}-\frac{p}{\rho}
\left(1+\phi\frac{\partial}{\partial \phi}\ln p^*\right)
\nabla_{\mathbf{V}}f^{(0)})-\left(1+\frac{1}{2}\phi\frac{\partial}{\partial \phi}\ln \chi\right)
\boldsymbol{\mathcal{K}}\left[f^{(0)}\right],  \label{8}
\end{equation}
\begin{equation}
\label{9} C_{ij}\left(
\mathbf{V}\right)=V_i\frac{\partial}{\partial V_j}f^{(0)}+\mathcal{K}_i\left[\frac{\partial}{\partial V_j}f^{(0)}
\right],
\end{equation}
\begin{equation}
D=\frac{1}{d}\nabla _{\mathbf{V}}\cdot\left(\mathbf{V}
f^{(0)}\right)-\frac{1}{2} \left( \zeta _{U}+\frac{2}{d} p^*\right)
\nabla _{\mathbf{V}}\cdot \left( \mathbf{V}
f^{(0)}\right)+\frac{1}{d}\mathcal{K}_{i}\left[\partial
_{V_{i}}f^{(0)}\right].   \label{10}
\end{equation}
Here, $\nabla_{\textbf{V}}\equiv \partial/\partial \textbf{V}$,
\begin{equation}
\label{11}
p^*\equiv\frac{p}{nT}=1+2^{d-2}(1+\alpha)\chi \phi,
\end{equation}
and $\mathcal{K}_{i}$ is the operator
\begin{equation}
\mathcal{K}_{i}[X] =\sigma^{d}\chi(\phi)\int \dd \mathbf{v}_{2}\int \dd\widehat{\boldsymbol {\sigma
}}\Theta (\widehat{\boldsymbol {\sigma}} \cdot
\mathbf{g})(\widehat{\boldsymbol {\sigma }}\cdot
\mathbf{g})
\widehat{\sigma}_i\left[ \alpha
^{-2}f^{(0)}(\mathbf{v} _{1}^{\prime
\prime})X(\mathbf{v}_{2}^{\prime \prime
})+f^{(0)}(\mathbf{v}_{1})X(\mathbf{v}_{2})\right],  \label{12}
\end{equation}
where ${\bf v}_1^{\prime\prime}={\bf v}_1-\frac{1}{2}(1+\alpha^{-1})(\widehat{\boldsymbol {\sigma }}\cdot
\mathbf{g})\widehat{\boldsymbol {\sigma }}$ and ${\bf v}_2''={\bf v}_2+\frac{1}{2}(1+\alpha^{-1})(\widehat{\boldsymbol {\sigma }}\cdot\mathbf{g})\widehat{\boldsymbol {\sigma }}$. In Eq.\ \eqref{10}, $\zeta_U$ is defined through the expression
\begin{equation}
\label{12.1}
\zeta=\zeta^{(0)}+\zeta_U\nabla \cdot {\bf U},
\end{equation}
where $\zeta^{(0)}$ is the cooling rate evaluated at zeroth-order.

With the distribution function $f^{(1)}$ determined, the goal now is to evaluate the heat and momentum fluxes as well as the cooling rate to first order in the spatial gradients. From these constitutive equations one can identify the forms of the Navier-Stokes transport coefficients.

\section{Pressure tensor}

To first order in the spatial gradients, the pressure tensor is given by
\begin{equation}
\label{13}
P_{ij}^{(1)}=-\eta\left( \partial _{i}U_{j}+\partial _{j}U_{i}-\frac{2}{d}\delta _{ij}\nabla \cdot
\mathbf{U} \right) -\gamma  \nabla \cdot \mathbf{U} \delta_{ij},
\end{equation}
where $\eta$ is the shear viscosity and $\gamma$ is the bulk viscosity. Their expressions are given by
\begin{equation}
\label{14}
\eta=\eta_k\left[1+\frac{B_2}{d+2}n^*\chi (1+\alpha)\right]+\frac{d}{d+2}\gamma,
\end{equation}
\begin{equation}
\label{15}
\gamma=B_3\frac{d+1}{4d^2}m\sigma^{d+1}\chi(1+\alpha)n^2v_0 I_\gamma(\alpha),
\end{equation}
where $n^*=n\sigma^d$. The subscript $k$ denotes the contributions from the kinetic part of the pressure tensor. The kinetic part $\eta_k$ to the shear viscosity is expressed in terms of the solution to the integral equation (\ref{4}) as
\begin{equation}
\label{16}
\eta_k=-\frac{1}{(d-1)(d+2)}\int\; \dd{\bf v} D_{ij} \mathcal{C}_{ij}({\bf V}),
\end{equation}
where $D_{ij}=m(V_iV_j-\frac{1}{d}V^2\delta_{ij})$. The dimensionless integral $I_\gamma$ is
\begin{equation}
\label{17}
I_\gamma=\frac{1}{n^2v_0}\int \dd{\bf V}_1\int \dd{\bf V}_2 f^{(0)}({\bf V}_1)f^{(0)}({\bf V}_2)g,
\end{equation}
where $v_0=\sqrt{2T/m}$ is the thermal velocity, $g=|\mathbf{V}_1-\mathbf{V}_2|$, and the coefficients $B_n$ are defined by
\begin{equation}
B_{n}\equiv \pi ^{\left( d-1\right) /2}\frac{\Gamma \left(
\frac{n+1}{2} \right) }{\Gamma \left( \frac{n+d}{2}\right) }.
\label{18}
\end{equation}

The kinetic part $\eta_k$ follows directly from the integral equation (\ref{4}) by multiplying it by $D_{ij}$ and integrating over the velocity. The final result can be written in the form $\eta_k=\eta_0\eta_k^*$, where
\begin{equation}
\label{20}
\eta_k^*=\left(\nu_\eta^*-\frac{1}{2}\zeta_0^*\right)^{-1}\left[1-\frac{2^{d-2}}{d+2}(1+\alpha)
(1-3 \alpha)\phi \chi \right],
\end{equation}
and
\begin{equation}
\label{19}
\eta_0=\frac{(d+2)\pi}{4\Omega_d}\sigma^{1-d}\sqrt{\frac{mT}{\pi}}
\end{equation}
is the low density value of the shear viscosity in the elastic limit. In these expressions, $\zeta_0^*\equiv\zeta^{(0)}/\nu_0$ and I have introduced the (reduced) collision frequency
\begin{equation}
\label{21}
\nu_\eta^*=\frac{\int \dd{\bf v} D_{ij}({\bf V}){\cal L}{\cal C}_{ij}({\bf V})}
{\nu_0\int \dd{\bf v}D_{ij}({\bf V}){\cal C}_{ij}({\bf V})},
\end{equation}
where $\nu_0=nT/\eta_0$. Upon deriving Eq.\ (\ref{20}), use has been made of the result
\begin{equation}
\label{22}
\int\; \dd{\bf V}\; D_{ij}({\bf V}) \mathcal{K}_i\left[\frac{\partial}{\partial V_j}f^{(0)}
\right]=2^{d-2}(d-1)nT\chi \phi (1+\alpha)(1-3\alpha).
\end{equation}

It is worthwhile remarking that, so far the expressions \eqref{14} and \eqref{20} for $\eta$ and \eqref{15} for $\gamma$ are \emph{exact}. However, in order to get the dependence of both transport coefficients on the coefficient of restitution $\alpha$ and the density $\phi$, one needs to know the explicit form of the reference distribution function $f^{(0)}$ and the solution $\mathcal{C}_{ij}$ to the integral equation \eqref{4}. The determination of $f^{(0)}$ to leading order in the Sonine polynomial expansion yields \cite{NE98}
\begin{equation}
\label{22.1}
f^{(0)}(V)\to f_M(V)\left\{1+\frac{c}{4}\left[\left(\frac{mV^2}{2T}\right)^2-\frac{d+2}{2}\frac{mV^2}{T}
+\frac{d(d+2)}{4}\right]\right\},
\end{equation}
where
\begin{equation}
\label{22.2}
f_M(V)=n\left(\frac{m}{2\pi T}\right)^{d/2}e^{-mV^2/2T}
\end{equation}
is the local equilibrium distribution function and \cite{NE98}
\begin{equation}
\label{22.3}
c=\frac{32(1-\alpha)(1-2\alpha^2)}{9+24d-\alpha(41-8d)+30(1-\alpha)\alpha^2}.
\end{equation}
Moreover, considering the approximation \eqref{22.1} for $f^{(0)}$ and neglecting nonlinear terms in $c$, the integral $I_\gamma$ is
\begin{equation}
\label{25}
I_\gamma=\sqrt{2}\frac{\Gamma\left(\frac{d+1}{2}\right)}{\Gamma\left(\frac{d}{2}\right)}\left(1-\frac{c}{32}\right).
\end{equation}

With respect to $\mathcal{C}_{ij}$, I'll consider a modified version \cite{GSM07} of the Sonine polynomial expansion where the distribution $f^{(0)}(V)$ is taking as the weight function in the Sonine expansion instead of the simpler Maxwellian form $f_M(V)$. Thus, in the case of the pressure tensor, the leading Sonine approximation to $\mathcal{C}_{ij}(\textbf{V})$ is
\begin{equation}
\label{23}
\mathcal{C}_{ij}({\bf V})\to -f^{(0)}(V)\frac{1}{nT^2}\frac{\eta_k}{1+\frac{c}{2}}D_{ij}({\bf V}).
\end{equation}
The collision integral $\nu_\eta^*$ can be evaluated when one considers the form \eqref{23} for $\mathcal{C}_{ij}(\textbf{V})$. Neglecting nonlinear terms in $c$ the result is \cite{GSM07}
\begin{equation}
\label{24}
\nu_\eta^*=\frac{3}{4d}\chi \left(1-\alpha+\frac{2}{3}d\right)(1+\alpha)
\left(1+\frac{7}{32}c\right).
\end{equation}

\section{Heat Flux}

The constitutive form for the heat flux in the Navier-Stokes approximation is
\begin{equation}
\label{26}
{\bf q}^{(1)}=-\kappa \nabla T-\mu \nabla n,
\end{equation}
where $\kappa$ is the thermal conductivity, and $\mu$ is an additional transport coefficient not
present in the elastic case. After some algebra, both transport coefficients can be written as
\begin{equation}
\label{27}
\kappa=\kappa_k\left[1+3\frac{2^{d-2}}{d+2}\phi \chi (1+\alpha)+\frac{B_3}{8d}\frac{m\sigma^{d+1}\chi}{T}(1+\alpha)n^2v_0^3I_\kappa\right],
\end{equation}
\begin{equation}
\label{28}
\mu=\mu_k\left[1+3\frac{2^{d-2}}{d+2}\phi \chi (1+\alpha)\right],
\end{equation}
where the dimensionless integral $I_\kappa$ is given by
\begin{equation}
\label{29}
I_\kappa=\frac{1}{n^2v_0^3}\int \dd{\bf V}_1\int \dd{\bf V}_2 f^{(0)}({\bf V}_1)f^{(0)}({\bf V}_2)\left[g^{-1}({\bf g}\cdot {\bf G})^2+gG^2+\frac{3}{2}g({\bf g}\cdot {\bf G})+\frac{1}{4}g^3\right],
\end{equation}
and $\mathbf{G}=\frac{1}{2}(\mathbf{V}_1+\mathbf{V}_2)$ is the velocity of center of mass.

The kinetic parts $\kappa_k$ and $\mu_k$ are defined as
\begin{equation}
\label{30}
\kappa=-\frac{1}{dT}\int\, \dd{\bf v} {\bf S}({\bf V})\cdot {\boldsymbol {\cal
A}}({\bf V}),
\end{equation}
\begin{equation}
\label{31}
\mu=-\frac{1}{dn}\int\, \dd{\bf v} {\bf S}({\bf V})\cdot {\boldsymbol {\cal
B}}({\bf V}),
\end{equation}
where
\begin{equation}
\label{32}
{\bf S}({\bf V})=\left(\frac{m}{2}V^2-\frac{d+2}{2}T\right){\bf V}.
\end{equation}

The kinetic part of the thermal conductivity is obtained by multiplication of Eq.\ (\ref{2}) by ${\bf S}({\bf V})$ and integration over the velocity. The result is
\begin{eqnarray}
\label{33}
\kappa_k &=& -\frac{1}{dT\nu_0}\left(\nu_\kappa^*-2\zeta_0^{*}\right)^{-1}\int \dd{\bf V} {\bf S}({\bf V})\cdot
{\bf A}({\bf V}) \nonumber\\
&=&\frac{1}{dT\nu_0}\left(\nu_\kappa^*-2\zeta_0^{*}\right)^{-1}\left\{
\frac{d(d+2)}{2m}nT^2 (1+c)-\frac{1}{2}\int\; \dd{\bf V}{\bf S}({\bf V})\cdot
\boldsymbol{\mathcal{K}}\left[\nabla_{\mathbf{V}}\cdot \left( \mathbf{V}
f^{(0)}\right) \right]\right\},
\end{eqnarray}
where
\begin{equation}
\label{34}
\nu_\kappa^*=\frac{\int \dd{\bf v} {\bf S}({\bf V})\cdot {\cal L}\boldsymbol{\mathcal{A}}({\bf V})}
{\nu_0\int \dd{\bf v}{\bf S}({\bf V})\boldsymbol{\mathcal{A}}({\bf V})}.
\end{equation}
The last term on the right hand side of Eq.\ (\ref{33}) can be evaluated more explicitly and the result is
\begin{equation}
\label{34}
\int\; \dd{\bf V}{\bf S}\cdot
\boldsymbol{\mathcal{K}}\left[\nabla_{\mathbf{V}}\cdot \left( \mathbf{V}
f^{(0)}\right) \right]=-\frac{3}{8}\Omega_d \frac{n^2T^2}{m}\sigma^d \chi(1+\alpha)^2\left[2\alpha-1+\frac{c}{2}(1+\alpha)\right].
\end{equation}
With this result, the kinetic part $\kappa_k$ is given by
\begin{equation}
\label{35}
\kappa_k=\kappa_0\frac{d-1}{d}\left(\nu_\kappa^*-2\zeta_0^{*}\right)^{-1}\left\{1+c+3\frac{2^{d-3}}{d+2}\phi \chi(1+\alpha)^2\left[2\alpha-1+\frac{c}{2}(1+\alpha)\right]\right\},
\end{equation}
where
\begin{equation}
\label{36}
\kappa_0=\frac{d(d+2)}{2(d-1)}\frac{\eta_0}{m}
\end{equation}
is the low density value of the thermal conductivity of an elastic gas.

To evaluate $\mu_k$, I multiply Eq.\ (\ref{3}) by ${\bf S}({\bf V})$ and integrate over
the velocity to get
\begin{equation}
\label{38} \mu_k=-\frac{2}{dn\nu_0}\left(2\nu_\mu^*-3\zeta_0^*\right)^{-1}\int\;\dd{\bf
V}{\bf S}({\bf V})\cdot \left[\left(1+\phi\partial_\phi\ln \chi\right)\zeta^{(0)}
\boldsymbol{\mathcal{A}}+{\bf B}\right],
\end{equation}
where
\begin{equation}
\label{38bis} \nu_\mu^*=\frac{\int \dd{\bf v} {\bf S}({\bf V})\cdot{\cal L}
{\bm{\mathcal{B}}}({\bf V})} {\nu_0\int \dd{\bf v}{\bf S}({\bf
V})\cdot{\bm{\mathcal{B}}}({\bf V})}.
\end{equation}
The last integral in Eq.\ (\ref{38}) is
\begin{equation}
\label{39} \int\;\dd{\bf V}{\bf S}\cdot {\bf B}=-\frac{d(d+2)}{4}\frac{nT^2}{m}c-
\left(1+\frac{1}{2}\phi\partial_\phi\ln \chi\right)\int\;\dd{\bf V}{\bf S} \cdot
\boldsymbol{\mathcal{K}}\left[f^{(0)}\right],
\end{equation}
where the last term can be evaluated in a similar way to Eqs.\ \eqref{22} and \eqref{34}. After some algebra, one gets
\begin{equation}
\label{40} \int\;\dd{\bf V}{\bf S} \cdot
\boldsymbol{\mathcal{K}}\left[f^{(0)}\right]=\frac{3}{8}\Omega_d\frac{n^*\chi
nT^2}{m}(1+\alpha)\left[\alpha(\alpha-1)+\frac{c}{12}(10+2d-3\alpha+3\alpha^2)\right].
\end{equation}
With these results, $\mu_k$ can be written as $\mu_k=(\kappa_0T/n)\mu_k^*$ where
\begin{eqnarray}
\label{41} \mu_k^*&=&2\left(2\nu_\mu^*-3\zeta_0^*\right)^{-1}\left\{\zeta_0^*\kappa_k^*
\left(1+\phi\partial_\phi\ln
\chi\right)+\frac{d-1}{2d}c+3\frac{2^{d-2}(d-1)}{d(d+2)}\phi \chi
(1+\alpha)\right. \nonumber\\
& &\left. \times \left(1+\frac{1}{2}\phi\partial_\phi\ln
\chi\right)\left[\alpha(\alpha-1)+\frac{c}{12}(10+2d-3\alpha+3\alpha^2)\right]\right\}.
\end{eqnarray}

The expressions \eqref{27} and \eqref{35} for the thermal conductivity $\kappa$ as well as Eqs.\ \eqref{28} and \eqref{41} for the coefficient $\mu$ are still exact. On the other hand, as in the case of the pressure tensor, to get the explicit forms of $\kappa$ and $\mu$ one has to determine the collision
frequencies $\nu_\kappa^*$ and $\nu_\mu^*$ and the integral $I_\kappa$. As before, the latter integral
can be evaluated by taking the leading Sonine approximation \eqref{22.1} for $f^{(0)}$ and
neglecting nonlinear terms in $c$. The result is
\begin{equation}
\label{42}
I_\kappa=\sqrt{2}\frac{\Gamma\left(\frac{d+3}{2}\right)}{\Gamma\left(\frac{d}{2}\right)}
\left(1+\frac{7}{32}c\right).
\end{equation}
With respect to $\nu_\kappa^*$ and $\nu_\mu^*$, one takes now the approximations
\begin{equation}
\label{43} \boldsymbol{\mathcal{A}}\to
-\frac{2}{d+2}\frac{1}{1+\frac{d+8}{16}c}\frac{m}{nT^2}\kappa_k \overline{{\bf S}}({\bf
V}),\quad \boldsymbol{\mathcal{B}}\to
-\frac{2}{d+2}\frac{1}{1+\frac{d+8}{16}c}\frac{m}{nT^2}\mu_k \overline{{\bf S}}({\bf
V}),
\end{equation}
where I have introduced the \emph{modified} Sonine polynomial
\begin{equation}
\label{44} \overline{{\bf S}}({\bf V})={\bf S}({\bf V})-\frac{d+2}{4}T c {\bf V}.
\end{equation}
Note that upon writing the approximations (\ref{43}), I have neglected nonlinear terms in $c$ as well as terms
proportional to the sixth cumulant of $f^{(0)}$. The collision integrals $\nu_\kappa^*$ and $\nu_\mu^*$ can be evaluated with the forms (\ref{43}) and the result is
\begin{equation}
\label{45}
\nu_\kappa^*=\nu_\mu^*=\frac{1+\alpha}{d}\chi\left[\frac{d-1}{2}+\frac{3}{16}(d+8)(1-\alpha)+\frac{296
+217d-3(160+11d)\alpha}{512}c\right].
\end{equation}

\section{Cooling rate}

The cooling rate $\zeta$ is given by
\begin{equation}
\label{46} \zeta=\zeta^{(0)}+\zeta_U \nabla \cdot {\bf U},
\end{equation}
where
\begin{equation}
\label{47} \zeta^{(0)}=\frac{d+2}{4d}(1-\alpha^2)\chi \left(1+\frac{3}{32}c\right)\nu_0.
\end{equation}
At first order in gradients, the proportionality constant $\zeta_U$ is a new transport
coefficient for granular fluids. This coefficient is given by
\begin{equation}
\label{48} \zeta_U=\zeta_{10}+\zeta_{11},
\end{equation}
where
\begin{equation}
\label{49} \zeta_{10}= -\frac{3}{4d^2}\Omega_d (1-\alpha^2)n^*\chi=-3\frac{2^{d-2}}{d}\chi \phi (1-\alpha^2),
\end{equation}
\begin{equation}
\zeta_{11}=\frac{1}{2nT}\frac{\pi ^{(d-1)/2}}{d\Gamma \left( \frac{d+3}{2} \right)}
\sigma ^{d-1}\chi m (1-\alpha^{2})\int d\mathbf{V}_{1}\,\int \dd
\mathbf{V}_{2}\,g^{3}f^{(0)}(\mathbf{V}_{1})\mathcal{D}(\mathbf{V}_{2}), \label{50}
\end{equation}
where the unknown functions $\mathcal{D}(\mathbf{V})$ are the solutions to the linear
integral equation (\ref{5}).

An approximate solution to the above integral equation (\ref{5}) can be obtained by
taking the Sonine approximation
\begin{equation}
\mathcal{D}(\mathbf{V})\rightarrow e_{D}f_{M}(\mathbf{V})F( \mathbf{V}), \label{51}
\end{equation}
where
\begin{equation}
F(\mathbf{V})=\left( \frac{m}{2T}\right) ^{2}V^{4}-\frac{d+2}{2}
\frac{m}{T}V^{2}+\frac{d(d+2)}{4}.  \label{52}
\end{equation}
The coefficient $e_D$ is given by
\begin{equation}
e_{D}=\frac{2}{d(d+2)}\frac{1}{n}\int \;\dd\mathbf{V}\;\mathcal{D}(
\mathbf{V})F(\mathbf{V}). \label{53}
\end{equation}
Substitution of Eq.\ (\ref{51}) into Eq.\ (\ref{50}) gives
\begin{equation}
\label{53bis}
\zeta_{11}=\frac{3(d+2)}{32d}\chi (1-\alpha^2)
\left(1+\frac{3}{64}c\right)\nu_0 e_D.
\end{equation}

The coefficient $e_D$ is determined by substituting Eq.\ (\ref{51}) into the integral
equation (\ref{5}), multiplying by $F({\bf V})$ and integrating over ${\bf V}$. The
result is
\begin{equation}
\label{54}
e_D=\left(\nu_\gamma^*-\frac{3}{2}\zeta_0^*\right)^{-1}\frac{2}{d(d+2)}\frac{1}{n\nu_0}\int\;
\dd{\bf V} F({\bf V}) D({\bf V}),
\end{equation}
where the term $\zeta_{11}c$ has been been neglected in accord with the present
approximation. Moreover, the terms proportional to $c$ coming from $\nu_\gamma^*$ and
$\zeta_0^*$ must be also neglected by consistency. In Eq.\ (\ref{54}), I have introduced
the collision frequency
\begin{equation}
\label{55} \nu_\gamma^*=\frac{\int \dd{\bf V} F({\bf V}){\cal L}[f_M(V) F({\bf V})]}
{\nu_0\int \dd{\bf V}f_M(V)F({\bf V})F({\bf V})}.
\end{equation}
The last integral in Eq.\ (\ref{54}) is given by
\begin{equation}
\label{56} \frac{2}{d(d+2)}\frac{1}{n}\int\;
\dd{\bf V} F({\bf V}) D({\bf V})=-\frac{3}{8d^2}\Omega_d n^*\chi (1+\alpha)\left(\frac{1}{3}-\alpha\right)c+
\frac{2}{nd^2(d+2)}\int \dd{\bf V} F \mathcal{K}_{i}\left[\partial
_{V_{i}}f^{(0)}\right],
\end{equation}
where
\begin{eqnarray}
\label{57}
\int \dd{\bf V} F \mathcal{K}_{i}\left[\partial
_{V_{i}}f^{(0)}\right]&=&\frac{3}{32}\Omega_d n^* n \chi (1+\alpha)\left\{
(1-\alpha^2)(5\alpha-1)-\frac{c}{12}\left[15\alpha^3-3\alpha^2+3(4d+15)\alpha-(20d+1)\right]\right\}\nonumber\\
&=&\frac{3}{32}\Omega_d n^* n \chi \omega^*.
\end{eqnarray}
The last equality of Eq.\ \eqref{57} defines the (reduced) quantity $\omega^*$. Moreover, the (reduced) collision frequency $\nu_\gamma^*$ is given by
\begin{equation}
\label{58}
\nu_\gamma^*=-\frac{1+\alpha}{192}\chi \left[30\alpha^3-30\alpha^2+(105+24 d) \alpha-56d-73\right].
\end{equation}

The expression for the coefficient $e_D$ can be obtained when one takes into account Eqs.\ (\ref{56}), (\ref{57})  and (\ref{58}). The result is
\begin{equation}
\label{59}
e_D=\nu_0^{-1}\left(\frac{d}{3}\nu_\gamma^*-\frac{d}{2}\zeta_0^*\right)^{-1}2^{d-3}
\phi\chi\left[\frac{\omega^*}{2(d+2)}-(1+\alpha)\left(\frac{1}{3}-\alpha\right)c\right].
\end{equation}
Therefore, the final expression for $\zeta_{U}$ can be written as
\begin{equation}
\label{60}
\zeta_U=3\frac{2^{d-2}}{d}\phi \chi (1-\alpha^2)\left(\frac{d+2}{64}\frac{\frac{\omega^*}{2(d+2)}-(1+\alpha)
\left(\frac{1}{3}-\alpha\right)c}{\frac{d}{3\chi}\nu_\gamma^*-
\frac{d+2}{8}(1-\alpha^2)}-1\right).
\end{equation}

\section{Summary of the main results}

In this section I provide a summary of the results obtained in this note. First, the reduced hydrostatic pressure $p^*$ is
\begin{equation}
\label{a1}
p^*=1+2^{d-2}(1+\alpha)\phi \chi.
\end{equation}
The (reduced) transport coefficients $\eta^*=\eta/\eta_0$, $\gamma^*=\gamma/\eta_0$, $\kappa^*=\kappa/\kappa_0$, and $\mu^*=(n\mu/T\kappa_0)$ are given, respectively by
\begin{equation}
\label{a2}
\eta^*=\eta_k^*\left[1+\frac{2^{d-1}}{d+2}\phi \chi (1+\alpha)\right]+\frac{d}{d+2}\gamma^*,
\end{equation}
\begin{equation}
\label{a3}
\gamma^*=\frac{2^{2d+1}}{\pi(d+2)}\phi^2 \chi (1+\alpha)\left(1-\frac{c}{32} \right),
\end{equation}
\begin{equation}
\label{a4}
\kappa^*=\kappa_k^*\left[1+3\frac{2^{d-2}}{d+2}\phi \chi (1+\alpha)\right]+\frac{2^{2d+1}(d-1)}{(d+2)^2\pi}
\phi^2 \chi (1+\alpha)\left(1+\frac{7}{32} c \right),
\end{equation}
\begin{equation}
\label{a5}
\mu^*=\mu_k^*\left[1+3\frac{2^{d-2}}{d+2}\phi \chi (1+\alpha)\right].
\end{equation}
The kinetic parts are
\begin{equation}
\label{a6}
\eta_k^*=\left(\nu_\eta^*-\frac{1}{2}\zeta_0^*\right)^{-1}\left[1-\frac{2^{d-2}}{d+2}(1+\alpha)
(1-3 \alpha)\phi \chi \right],
\end{equation}
\begin{equation}
\label{a7}
\kappa_k^*=\frac{d-1}{d}\left(\nu_\kappa^*-2\zeta_0^{*}\right)^{-1}\left\{1+c+3\frac{2^{d-3}}{d+2}\phi \chi(1+\alpha)^2\left[2\alpha-1+\frac{c}{2}(1+\alpha)\right]\right\},
\end{equation}
\begin{eqnarray}
\label{a8} \mu_k^*&=&\left(\nu_\kappa^*-\frac{3}{2}\zeta_0^*\right)^{-1}\left\{\zeta_0^*\kappa_k^*
\left(1+\phi\partial_\phi\ln
\chi\right)+\frac{d-1}{2d}c+3\frac{2^{d-2}(d-1)}{d(d+2)}\phi \chi
(1+\alpha)\right. \nonumber\\
& &\left. \times \left(1+\frac{1}{2}\phi\partial_\phi\ln
\chi\right)\left[\alpha(\alpha-1)+\frac{c}{12}(10+2d-3\alpha+3\alpha^2)\right]\right\}.
\end{eqnarray}
In these expressions, $c$, $\zeta_0^*$, $\nu_{\eta}^*$, and $\nu_{\kappa}^*$ are functions of
$\alpha$ and $\phi$. They are given by
\begin{equation}
\label{a9}
c(\alpha)=\frac{32(1-\alpha)(1-2\alpha^2)}{25+24d-\alpha(57-8d)-2(1-\alpha)\alpha^2},
\end{equation}
\begin{equation}
\label{a10}
\zeta_0^*=\frac{d+2}{4d}(1-\alpha^2)\chi \left(1+\frac{3}{32}c\right),
\end{equation}
\begin{equation}
\label{a11}
\nu_\eta^*=\frac{3}{4d}\chi \left(1-\alpha+\frac{2}{3}d\right)(1+\alpha)
\left(1+\frac{7}{32}c\right),
\end{equation}
\begin{equation}
\label{a12}
\nu_\kappa^*=\frac{1+\alpha}{d}\chi\left[\frac{d-1}{2}+\frac{3}{16}(d+8)(1-\alpha)+\frac{296
+217d-3(160+11d)\alpha}{512}c\right].
\end{equation}

The (reduced) cooling rate $\zeta^*\equiv \zeta/\nu_0$ can be written as
\begin{equation}
\label{a12.1} \zeta^*=\zeta_0^*+\zeta_U \nu_0^{-1}\nabla \cdot {\bf U},
\end{equation}
where
\begin{equation}
\label{a12.2} \zeta_0^{*}=\frac{d+2}{4d}(1-\alpha^2)\chi \left(1+\frac{3}{32}c\right),
\end{equation}
\begin{equation}
\label{a13}
\zeta_U=3\frac{2^{d-2}}{d}\phi \chi (1-\alpha^2)\left(\frac{d+2}{64}\frac{\frac{\omega^*}{2(d+2)}-(1+\alpha)
\left(\frac{1}{3}-\alpha\right)c}{\frac{d}{3\chi}\nu_\gamma^*-
\frac{d+2}{8}(1-\alpha^2)}-1\right),
\end{equation}
where
\begin{equation}
\label{a14}
\nu_\gamma^*=-\frac{1+\alpha}{192}\chi \left[30\alpha^3-30\alpha^2+(105+24 d) \alpha-56d-73\right],
\end{equation}
and
\begin{equation}
\label{a15}
\omega^*=(1+\alpha)\left\{
(1-\alpha^2)(5\alpha-1)-\frac{c}{12}\left[15\alpha^3-3\alpha^2+3(4d+15)\alpha-(20d+1)\right]\right\}.
\end{equation}

Aside from the terms explicitly neglected in Ref.\ \cite{L05}, all the above expressions are consistent with those previously obtained by J. F. Lutsko \cite{L05}, except in the coefficient of $\nu_\gamma^*$ in the denominator of Eq.\ (\ref{a13}). Moreover, in the case of hard spheres ($d=3$), the expressions \eqref{a2}--\eqref{a8} for the Navier-Stokes transport coefficients and \eqref{a13}--\eqref{a15} for the cooling rate also agree with those obtained in Refs.\ \cite{GD99} and \cite{G05}, respectively.

\end{document}